\pdfoutput=1

\documentclass[pra,superscriptaddress,amsmath,amssymb,showpacs,twocolumn]{revtex4}

\usepackage{graphicx,bm,amsmath}
\usepackage{hyperref}
\usepackage{color}

\newcommand{\be}{\begin{equation}}
\newcommand{\ee}{\end{equation}}
\newcommand{\bea}{\begin{eqnarray}}
\newcommand{\eea}{\end{eqnarray}}
\newcommand{\la}{\lambda}

\begin{document}

\title{Random matrices and entanglement entropy of trapped Fermi gases}

\author{Pasquale Calabrese} 
\affiliation{SISSA and INFN, via Bonomea 265, 34136 Trieste, Italy }
\author{Pierre Le Doussal} 
\affiliation{CNRS-Laboratoire de Physique
Th{\'e}orique de l'Ecole Normale Sup{\'e}rieure, 24 rue Lhomond, 75231
Paris Cedex, France}
\author{ Satya N. Majumdar}
\affiliation{CNRS-Universit\'e Paris-Sud, LPTMS, UMR8626-B\^at 100,91405 Orsay Cedex, France}


\begin{abstract}

We exploit and clarify the use of random matrix theory for the calculation of the entanglement entropy of free Fermi gases.
We apply this method to obtain analytic predictions for R\'enyi entanglement entropies of a one-dimensional 
gas trapped by a harmonic potential in all the relevant scaling regimes. 
We confirm our findings with accurate numerical calculations obtained by means of 
an ingenious discretisation of the reduced correlation matrix.
 
\end{abstract}


\maketitle

\section{Introduction}
During the last decade entanglement became a very powerful tool for the 
study of many-body quantum systems especially for the identification of critical and 
topological phases of matter (see e.g. Refs. \cite{AFOV-08,ECP-10,rev-cc} as reviews).
In this respect the most studied quantity is surely the (von Neumann or R\'enyi) entanglement entropy.
In terms of the reduced density matrix $\rho_A={\rm Tr}_{\bar A} \rho$ of a subsystem $A$
($\bar A$ denotes the complement of $A$), the order-$q$ R\'enyi entropy is defined as 
\be
S_q=\frac1{1-q}\ln {\rm Tr} \rho_A^q\,,
\label{def:Sq}
\ee
that in the limit $q\to 1$ reduces to the most studied von Neumann entropy $S_1$.
The knowledge of the  R\'enyi entropies for arbitrary values of $q$ contains much more information than the sole $S_1$ 
since from them one can extract the full spectrum of $\rho_A$ \cite{cl-08}.

From the definition and from the highly non-local character of Eq. (\ref{def:Sq}), it can appear extremely difficult to calculate the 
entanglement entropy even for the simpler models. 
However, a number of advanced analytic techniques have been developed in such a way to have a rather precise characterisation 
in many different classes of systems. 
These include one-dimensional conformal field theories \cite{holzhey,vidalent,CC-04}, 
spin-chains mappable to free fermions thanks to Toeplitz matrix techniques \cite{vidalent,JK-04,CCEN-10,ce-10,fc-11,ia-13},
higher dimensional lattice fermions with Widom conjecture \cite{Wolf-06}, holographic techniques \cite{holo}, renormalisation group \cite{mfs-09}, and many more. 
The entanglement entropies are also a crucial concept to understand the scaling and the working \cite{t-08}
of matrix product states algorithms \cite{mps}.

In this paper we discuss and develop the connection between the entanglement entropies and random 
matrix theory in free one-dimensional Fermi gases.
A similar connection was first highlighted in lattice models in Ref. \cite{km-05} and further developed in \cite{ce-10}. 
In two recent manuscripts \cite{MMSV14,e-14}, random matrix theory has been used to calculate the probability particle 
distribution (aka the full counting statistics) in a finite length interval, but not for the entanglement entropies. 
As we shall see, this approach  allows to clarify several concepts already present in the literature and provides 
also new results, such as the scaling of the entanglement entropy in a free fermion gas confined by a harmonic potential, 
a problem that so far has been studied only numerically \cite{v-12} and for which an analytical description was still missing.

The manuscript is organised as follows. 
In Sec. \ref{sec2} we briefly review the standard methods for the calculation of the entanglement entropy in Fermi gases 
and we establish the correspondence with random matrix theory. 
In Sec. \ref{sec3} we use this formalism to analytically calculate the entanglement entropies  
for a one dimensional Fermi gas trapped in an harmonic potential for  an interval symmetric with respect to the centre of the trap.
In the same section, we also confirm our findings by accurate numerical calculations. 
Finally in Sec. \ref{conl} we draw our conclusions and we discuss some possible generalisations and open issues. 
Some details about the density of eigenvalues of the overlap matrix have been relegated
to appendix \ref{appa}.

\section{Free fermion gases and Random matrix theory}
\label{sec2}

Let us consider a system of $N$ non-interacting spinless fermions with discrete one-particle energy spectrum.  
The many body wave functions $\Psi(x_1,...,x_N)$ is the Slater determinant 
built with the one-particle eigenstates, i.e. 
\be
\Psi(x_1,...,x_N)=\frac1{\sqrt{N!}}\det [\phi_k(x_n)], 
\ee
where the normalized wave functions $\phi_k(x)$ are the occupied single-particle energy levels. 
The ground state $\Psi_0(x_1,...,x_N)$ is obtained by filling the lowest $N$ energy levels.
The ground-state two-point correlation function is
\begin{equation}
C(x,y) \equiv \langle c^\dagger(x) c(y) \rangle = \sum_{k=1}^{N} \phi^*_k(x) \phi_k(y)\,, 
\label{cxy}
\end{equation} 
where $c(x)$ is the fermionic annihilation operator and the one-particle
eigenfunctions $\phi_k(x)$ are ordered according to their energies.
The Wick theorem allows to write the reduced density matrix of a spatial subsystem $A$ as \cite{pes}
\be
\rho_A \propto \exp \Big(- \int_{A} d y_1 dy_2 c^\dagger(y_1) {\cal H}(y_1,y_2) c(y_2)\Big)\,,
\label{rhoaa}
\ee
where ${\cal H}=\ln [(1-C)/C]$ and the normalization constant is fixed by requiring ${\rm Tr}\rho_A=1$.

\begin{figure}[t]
\includegraphics[width = \linewidth]{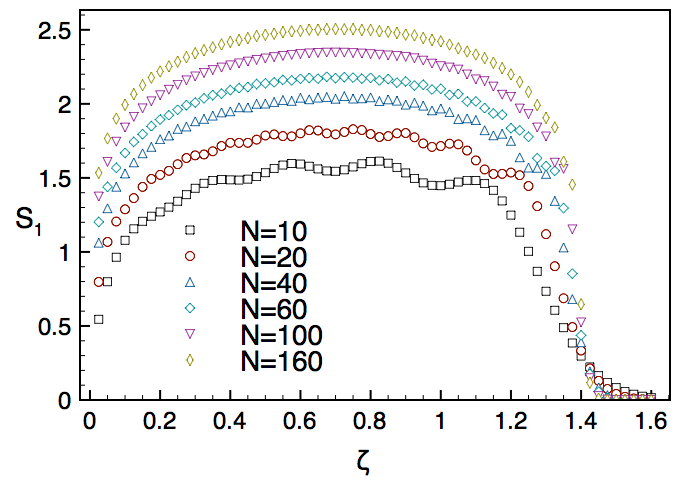}
\caption{(Color online) Entanglement entropy $S_1$ for a Fermi gas with $N$ particles trapped in a harmonic potential. 
We consider the bipartition in which the subsystem $A$ is the interval $A=[-\ell,\ell]$. 
We report the entanglement entropy as function of $\zeta=\ell/\sqrt{N}$ for different values of $N$.
The reported data are obtained from an ingenious discretisation of Eq. (\ref{entaC}). }
\label{fig:Sv}
\end{figure}

It is useful to define the correlation matrix restricted to the subsystem $A$ 
\be
C_A(x,y)\equiv I_A(x) C(x,y) I_A(y), 
\ee
with $I_A(x)$ being the characteristic function of the subsystem, i.e.
\be
I_A(x)=\begin{cases}
1 & x\in A ,\\ 
0& x\notin A. 
\end{cases}
\ee
A related quantity is the overlap matrix of the subsystem $A$ defined as \cite{CMV-11,CMV-11a}
\begin{equation}
{\mathbb A}_{nm} =  \int_{A} dz\, \phi_n^*(z) \phi_m(z),
\qquad n,m=1,...,N,
\label{aiodef}
\end{equation}
As shown in Refs. \cite{CMV-11,CMV-11a}, the overlap matrix and the restricted correlation matrix 
have the same spectrum although they act on different spaces. 
Using the quadratic form of the reduced density matrix (\ref{rhoaa}), the R\'enyi entanglement entropies 
can be written in terms of the overlap or correlation matrices as
\bea
S_q&=& \frac1{1-q} {\rm Tr} \ln [{\mathbb A}^q+(1-{\mathbb A})^q],\\
S_q&=&\frac1{1-q} {\rm Tr} \ln [C_A^q+(1-C_A)^q]. \label{entaC}
\eea
In terms of the eigenvalues $a_i$, common to the overlap and reduced correlation matrices,
the entanglement entropy is 
\be 
S_q=\sum_{i=1}^N e_q(a_i), \qquad
e_q(x)\equiv \frac1{1-q}  \ln [x^q+(1-x)^q].
\label{eq}
\ee
At this point there are two possible roads for a numerical evaluation of the entropy.
The first possibility is to explicitly construct the overlap matrix, 
find its eigenvalues numerically, and from them computing $S_q$. 
This  numerical approach has been effectively applied for the determination of the entanglement entropy of 
Fermi gases in many equilibrium \cite{CMV-11,CMV-11a,CMV-11b,CMV-12,v-12,ep-13,o-14,nv-13} and non-equilibrium situations \cite{nv-13,v-12b,csc-13,ckc-13},
as well as to the related statistics of particle number in the subsystem \cite{CMV-12b,si-12,lbb-12,er-13,e-14,MMSV14,c-14}
(we mention that the entanglement entropies of trapped lattices gases were numerically studied in \cite{cv-10}).
A second possibility is to extract the spectrum from the reduced correlation matrix. 
While at first this can sound awkward, because we should work with a continuous kernel, some very effective discretisations 
have been developed \cite{born}, which allow a much faster computation of the entropies especially when
the integrals defining the elements of the overlap matrix (\ref{aiodef}) can not be analytically performed.  
In Fig. \ref{fig:Sv} we report the numerically evaluated entanglement entropy $S_1$ for the model studied
in this paper which is a Fermi gas trapped in a harmonic potential. 
We only consider the case in which the subsystem is the symmetric interval $A=[-\ell,\ell]$.
We calculated the spectrum of $C_A$ by using the Gauss-Legendre discretisation proposed in Ref. \cite{born}.
We found that in order to achieve a precision of about $10^{-8}$ on the entropy, the discretised matrix should have a 
dimension growing linearly in $N$ which is the same as the overlap matrix, but its elements must not be calculated 
by numerical integration. 
The main reason for this high efficiency is that convergence is exponential in the  number of steps 
on Gauss-Legendre discretisation \cite{born}.
We checked for several values of $N$ that the spectrum of the reduced correlation matrix obtained in this way is the same 
as the one of the overlap matrix, but its numerical determination is much faster.
Obviously, every time that  the overlap matrix is analytically evaluable (as e.g. in the cases considered in \cite{CMV-11a}),
there is no advantage in this procedure and the overlap matrix method remains favourable. 
We mention that the results reported in Fig. \ref{fig:Sv} are equivalent to those already reported in Ref. \cite{v-12}.

\subsection{The connection with random matrix theory}

The connection with random matrix theory  \cite{MMSV14,e-14}
starts from the definition of the characteristic polynomial of ${\mathbb A}$ (or $C_A$) 
\be \label{DA} 
D_A(\la)=\prod_{i=1}^N (\la-a_i)=\det[\la {\mathbb I - \mathbb A}],
\ee
which is a standard tool in the analytic calculation of the entanglement entropy \cite{JK-04,ce-10}.
This characteristic polynomial $D_A(\la)$ can be straightforwardly written as a random matrix average. 
Indeed by definition we have (using the completeness of the eigenfunctions $\phi_m(z)$ on the full line)
\begin{multline}
D_A(\la)=\det\left[ \lambda \int_{-\infty}^\infty dz \phi_n^*(z) \phi_m(z) 
\right. \\\left. 
- \int_A dz \phi_n^*(z) \phi_m(z)
\right]\\
= \det\left[ \int_{-\infty}^\infty dz (\la -I_A(z))\phi_n^*(z) \phi_m(z)\right].
\end{multline}
At this point we can use the Cauchy-Binet identity 
\begin{multline}
\int dx_1\dots dx_N \det[f_i(x_j)]\det[g_k(x_l)]\prod_{i=1}^N h(x_i)=\\=
N! \det \left[\int dx h(x) f_i(x) g_j(x)\right]\,,
\end{multline}
to rewrite $D_A(\la)$ as 
\begin{multline}
D_A(\la)= \frac1{N!} \int dx_1 \dots dx_N \det[\phi_i(x_j)]\det[\phi_k(x_l)]\\ \times
\prod_{i=1}^N (\la-I_A(x_i))=\\=
\int dx_1 \dots dx_N |\Psi_0(x_1,\dots x_N)|^2 \prod_{i=1}^N (\la-I_A(x_i))\,,
\end{multline}
where we recognized $\Psi_0(x_1,...,x_N)={\rm det} [\phi_k(x_n)]/\sqrt{N!}$.
Thus, every time that  $|\Psi_0(x_1,\dots x_N)|^2$ corresponds to a random matrix average $\langle \cdot \rangle_{\rm RM}$, 
when the $x_i$ are related to eigenvalues of a random matrix (see below), the above equation is equivalent to 
\be
D_A(\la)=\Big\langle \prod_{i=1}^N (\la-I_A(x_i)) \Big\rangle_{\rm RM}.
\label{Da.1}
\ee
We will list and analyse in the following a number of interesting random matrix averages for 1D Fermi gases, but first  
we proceed to further simplifications and interpretation of the above average.
We will also remove the subscript ${\rm RM}$ from the averages.

To this aim, let us introduce the operator counting particle number in the subsystem $A$ (here 
$\hat n (x)=c^\dagger (x) c(x)$ is the particle density)
\be
N_A=\sum_{i=1}^N I_A (x_i) = \int_A \hat n (x) dx,  
\ee
and its generating function 
\be
\chi(s)\equiv \langle e^{-s N_A}\rangle= \Big\langle \prod_{i=1}^N e^{-s I_A(x_i)}\Big\rangle.
\ee
(Often  $\chi(is)$ is called generating function, but this is not important for what follows).
Since
\be
e^{-s I_A(x)}=\begin{cases}
e^{-s} & x\in A ,\\ 
1& x\notin A, 
\end{cases}
\ee
we have
\be
e^{-s I_A(x)}= e^{-s}  I_A(x)+(1-  I_A(x))=1-(1-e^{-s})  I_A(x),
\ee
and then 
\begin{multline}
\chi(s)=\Big\langle \prod_{i=1}^N [1- (1-e^{-s}) I_A (x_i)]\Big\rangle =\\
= (1-e^{-s})^N \Big\langle \prod_{i=1}^N \Big[\Big(\frac1{1-e^{-s}}\Big) -I_A(x_i)\Big]\Big\rangle\,.
\end{multline}
Setting
\be
\la=\frac1{1-e^{-s}}\Rightarrow e^{-s}= \frac{\la-1}\la\,,
\ee
we have 
\be
\Big\langle  \Big(\frac{\la-1}\la\Big)^{N_A}\Big\rangle=\frac1{\la^N} \Big\langle \prod_{i=1}^N (\la-I_A(x_i) )\Big\rangle.
\ee
Thus, plugging the above equation in Eq. (\ref{Da.1}), we have
\be
D_A(\la)= \la^N \chi\Big(e^{-s}=1-\frac1\la\Big)= \la^N \Big\langle  \Big(\frac{\la-1}\la\Big)^{N_A}\Big\rangle.
\label{DA.2}
\ee
This is a very compact expression for the characteristic polynomial valid for arbitrary number of particles $N$ and arbitrary random 
matrix average.
Although it appeared (in a more or less explicit form) a few times in the literature, its general validity has not been  appreciated enough. 

\begin{figure}[t]
\includegraphics[width = \linewidth]{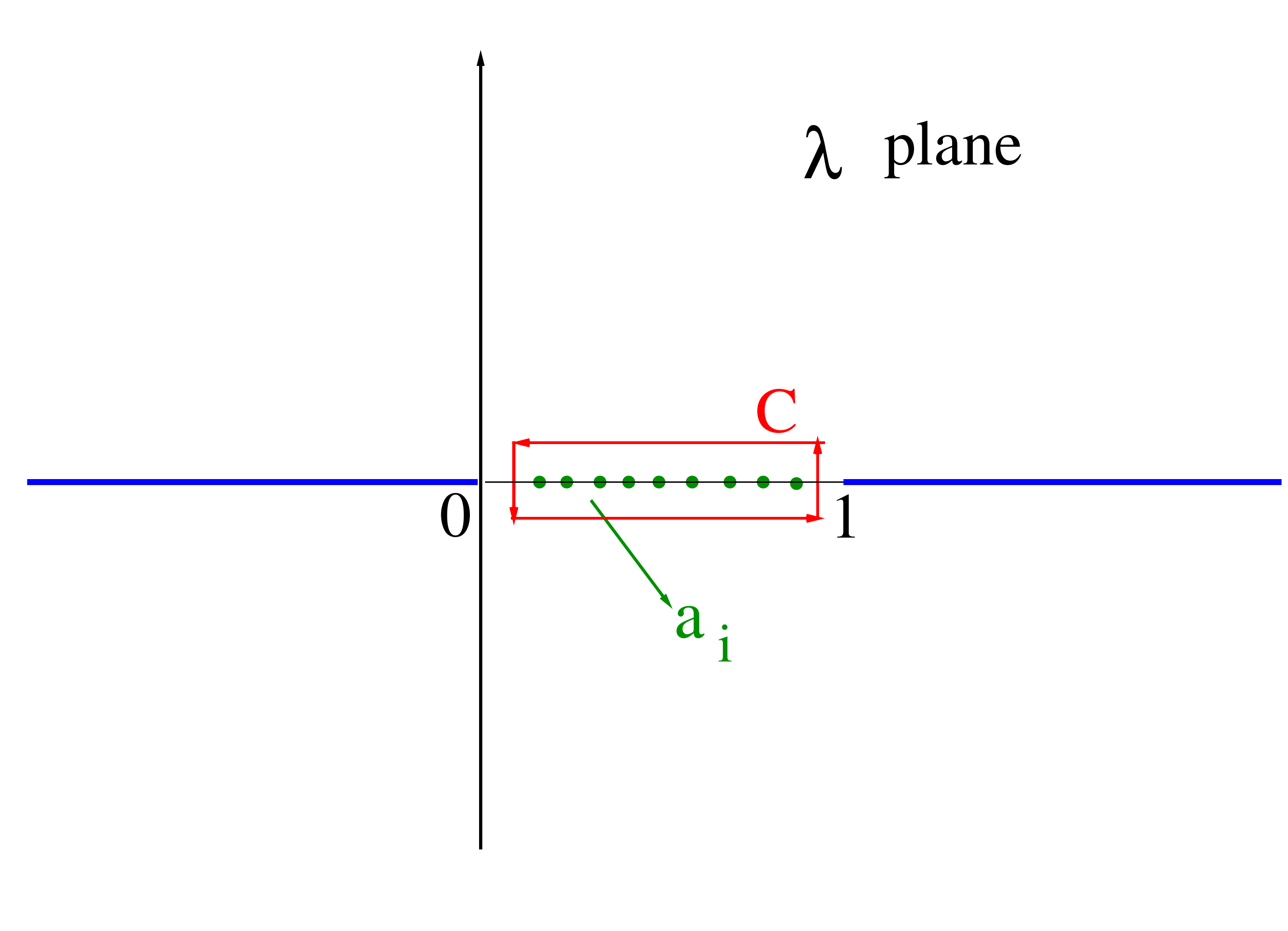}
\caption{(Color online) The rectangular counterclockwise contour $C$ in the 
complex $\la$ plane encloses the poles at $a_i$'s shown by dots. 
The left vertical line of $C$ is to right of $\la=0$ and
the right vertical line is to the left of $\la=1$. }
\label{fig:contour}
\end{figure}

In order to calculate the entropies let us introduce the resolvent 
\be
F(\lambda)=\sum_{i=1}^N \frac1{\la-a_i}={\rm Tr} \frac1{\la {\mathbb I-\mathbb A}}\,,
\label{fmu.1}
\ee
which is related to $D_A(\la)$ as 
\be
F(\la)=\frac{D'_A(\la)}{D_A(\la)}=\frac{d}{d\la} \ln D_A(\la)\,.
\label{Fla}
\ee
Using Eq. (\ref{DA.2}) for $D_A(\la)$ we have after simple algebra
\be
F(\la)=\frac{N}\la +\frac1{\la(\la-1)} 
\frac{\Big  \langle N_A \Big(1-\frac{1}\la\Big)^{N_A}\Big\rangle}{\Big\langle  \Big(1-\frac{1}\la\Big)^{N_A}\Big\rangle}.
\label{fmu.2}
\ee
Given that the entropies are given by Eq. (\ref{eq}), we immediately have 
\bea
S_q= \int_{C} \frac{d\la}{2\pi i}\,  e_q(\la)\, F(\la),
\label{intexp.1}
\eea
where the contour $C$ in the complex $\la$ plane goes counterclockwise over the rectangle: $[0,1]\times 
[-\epsilon, \epsilon]$, as shown in Fig. (\ref{fig:contour}), with
$\epsilon\to 0^+$ eventually.
Note that the function $e_q(\la)$ has branch cuts for ${\rm Re}(\la)<0$ and
${\rm Re}(\la)>1$ (this is equivalent to the analogous formulas for spin chains \cite{JK-04,ce-10}).
Upon substituting the definition of $F(\la)$ from Eq. (\ref{fmu.1})
on the right hand side of Eq. (\ref{intexp.1}) and 
calculating the residues around the
poles $0\le a_i\le 1$, gives this result immediately.

Plugging the expression of $F(\la)$ from Eq. (\ref{fmu.2}) into Eq.  (\ref{intexp.1}), one arrives to a rather compact exact expression 
for the entropy, valid for all $N$
\begin{multline}
S_q= \frac{1}{(1-q)}\, \frac{1}{2\pi i}\, \int_C 
\frac{d\la}{\la(\la-1)}\, \ln\left[\la^q+ (1-\la)^q\right]\\ \times 
\frac{\Big\langle
N_A\left(1-\frac{1}{\la}\right)^{N_A}\Big\rangle}{\Big\langle
\left(1-\frac{1}{\la}\right)^{N_A}\Big\rangle} \, .
\label{entropy.2}
\end{multline}
The term $N/\la$ in Eq. (\ref{Fla}) does not contribute to the entropy $S_q$ because, inside the integration contour, it provides 
an analytic function  with zero residue. 
By writing further, $1-1/\la= e^{-s}$, one can write a slightly more
compact expression for the ratio
\bea
\frac{\Big\langle
N_A\left(1-\frac{1}{\la}\right)^{N_A}\Big\rangle}{\Big\langle
\left(1-\frac{1}{\la}\right)^{N_A}\Big\rangle}= -\frac{\partial}{\partial s} 
\ln \left[\langle e^{-s N_A}\rangle\right]\, .
\label{compact.1}
\eea

Finally, these expressions allow   us to derive the asymptotic large $N$
density of eigenvalues $\rho(a)$ of the overlap matrix (or reduced correlation matrix) 
which is defined by the implicit relation
\be
N \int da \frac{\rho(a)}{\la-a}  = F(\la)\,,
\ee
leading to
\be
\rho(a) = -\frac{1}{\pi} \lim_{\epsilon\to0^+} {\rm Im} F(a+i \epsilon) .
\label{rhoa}
\ee
We discuss explicitly the density of eigenvalues $\rho(a)$ for a trapped Fermi gas in Appendix \ref{appa}.

\subsection{Gaussian distribution}

An immediate consequence of the exact formula in Eq. (\ref{entropy.2})
is the well-known relation \cite{CMV-12b,KRS-06,KL-09,SRL-10,SRFKL-11} between the variance of $N_A$ and entropies 
in the case the random variable $N_A$ is a {\em pure}
Gaussian with mean $\langle N_A\rangle$ and variance $V_{N_A}$, i.e.,
\bea
N_A= \langle N_A\rangle + \sqrt{V_{N_A}}\, {\cal N}(0,1),
\label{gaussian.1}
\eea
where ${\cal N}(0,1)$ is a standard normal Gaussian variable with zero  mean and unit variance. 
Indeed, using the Gaussian property of ${\cal N}(0,1)$,   it follows immediately that 
\bea
\langle e^{-s N_A}\rangle = e^{-s\, \langle N_A\rangle + \frac{s^2}{2}\, 
V_{N_A}}\,.
\label{gaussian.2}
\eea
Taking logarithm and deriving with respect to $s$ as in Eq. 
(\ref{compact.1}) we obtain 
\bea
\frac{\Big\langle
N_A\left(1-\frac{1}{\la}\right)^{N_A}\Big\rangle}{\Big\langle  
\left(1-\frac{1}{\la}\right)^{N_A}\Big\rangle}= \langle N_A\rangle + V_{N_A}\, 
\ln \left(1-\frac{1}{\la}\right) .
\label{compact.2}
\eea
Plugging this expression in Eq. (\ref{entropy.2}) gives
\begin{multline}
S_q=  \frac{1}{(1-q)}\, \frac{1}{2\pi i}\, \int_C
\frac{d\la}{\la(\la-1)}\, \ln\left[\la^q+ (1-\la)^q\right] \\ \times \left[\langle 
N_A\rangle + V_{N_A}\, \ln \left(1-\frac{1}{\la}\right)\right]\,.
\label{entropy.3}
\end{multline}
The contour integral with the constant term $\langle N_A\rangle$ vanishes
since the integrand in analytic inside the contour (which does not
include the poles at $\la=0$ and $\la=1$). 
This leaves us with 
\begin{multline}
S_q= \frac{V_{N_A}}{(1-q)}\, \frac{1}{2\pi\, i}\, \int_C
\frac{d\la}{\la(\la-1)}\\ \times \ln\left[\la^q+ (1-\la)^q\right]\,
\ln \left(1-\frac{1}{\la}\right),
\label{entropy.4}
\end{multline}
which is an exact expression for entropy when $N_A$ is a pure Gaussian.
The contour integral in Eq. (\ref{entropy.4}) can be performed exactly
in the limit $\epsilon\to 0^+$. The contributions from the vertical
portions vanish as $\epsilon\to 0^+$ and the contributions from the
horizontal portions gives a real integral over $\la\in [0,1]$ as follows
\begin{multline}
S_q= - \frac{V_{N_A}}{\pi (1-q)}\, \int_0^1 \frac{dx}{x(x-1)} \\ \times
\ln\left[x^q+ (1-x)^q\right]\, {\rm Im}\left[ \ln 
\Big(1-\frac{1}{x+i\epsilon}\Big)\right]_{\epsilon\to 0^+}\,.
\label{entropy.5}
\end{multline}
Using ${\rm Im}\left[ \ln (1-\frac{1}{x+i\epsilon})\right]_{\epsilon\to 0^+}= \pi$ then gives the
final result for the entropy, given that $N_A$ is a pure Gaussian,
\begin{multline}
S_q= - \frac{V_{N_A}}{(1-q)}\, \int_0^1 \frac{dx}{x(x-1)}\,
\ln\left[x^q+ (1-x)^q\right]\\ = \frac{\pi^2}{6} 
\left(1+\frac{1}{q}\right)\, V_{N_A}\, .
\label{entropy.6}
\end{multline}
Although this relation between entropy and fluctuations is well-known in the literature \cite{CMV-12b,SRFKL-11}, 
we find the above derivation very instructive from the random matrix point of view. 

\subsection{Examples of random matrices ensembles and corresponding fermionic systems}

For a Fermi gas in a ring of length $L$ with periodic boundary conditions, the normalized one-particle
wave-functions are plane waves $\phi_{k}(x)=e^{2\pi i k x/L}/\sqrt{L}$ and the corresponding 
many-body wave-function $\Psi_0$ gives the circular unitary ensemble (CUE). 
This random matrix ensemble has been already study in the context of the entanglement entropy of 
spin chains \cite{km-05,ce-10} and the these results have been exported to the Fermi gas in \cite{CMV-11b}.
In the case when $A$ is an interval of length $\ell$ embedded in a finite system of length $L$, the leading and subleading 
behavior for the entropy has been obtained  in \cite{CMV-11b}. 
The asymptotic large $N$ behavior of the entropies for fixed ratio $\ell/L$ and at finite density $n=N/L$ 
(obtained by means of the Fisher-Hartwig conjecture)
is  \cite{CMV-11,CMV-11b}
\be
S_q =\frac16\left(1+\frac1q\right)\ln\left (2N\sin\pi \frac\ell{L}\right)+ E_q+o(N^0), 
\label{FHres}
\ee 
and the constant $E_q$ is given by  \cite{JK-04}
\begin{multline}
E_q=
\left(1+\frac1q\right)\int_0^\infty \frac{dt}t 
\left[\frac{1}{1-q^{-2}}
\right. \\ \left. \times
\left(\frac1{q\sinh t/q}-\frac1{\sinh t}\right)
\frac1{\sinh t}-\frac{e^{-2t}}6\right]\, .
\label{cnp}
\end{multline}
Random  matrices techniques are instead a needed tool to access some of the  corrections to the above leading behavior, 
see Ref. \cite{ce-10}.
More general results for the case when $A$ is composed of disjoint intervals are also known \cite{CMV-11b}.

It is important to  mention that the functional dependence of the entanglement entropy \eqref{FHres} over $\ell$ and $L$
is a general prediction of conformal field theory \cite{CC-04,cc-rev}. Indeed from the well-known infinite system
result 
\be
S_q=\frac16\left(1+\frac1q\right)\ln \ell + E_q,
\ee
one obtains Eq. (\ref{FHres}) with the replacement $\ell\to N \sin \pi \ell/L$, as a consequence of the mapping from the plane 
to a cylinder of circumference $L$ \cite{CC-04}. 
This simple result is indeed valid for a general correlation function of primary operators 
(in CFT the entanglement entropies for integer $q$ are  correlation functions of the so called twist fields 
\cite{ccd-08,cc-rev}).
This is a very powerful prediction for the finite size-scaling function of the entanglement entropy for homogeneous systems 
whose analog in the presence of a harmonic potential  will be calculated  in the following section.

For a gas of spinless fermions confined in the interval $[0,L]$ by a hard-wall
potential, the one-particle wave functions are $\phi_k(x) =  {\sqrt{\frac2L}}\sin\left[{\pi}k\frac{x}L\right]$.
In this case the corresponding random matrix ensemble is $O^+(2N)$ symmetric \cite{km-05}, but the 
consequences of this correspondence have not been studied in great details yet. 
The asymptotic large $N$ behavior of the entanglement entropy has been obtained by using generalisation of the 
Fisher-Hartwig conjecture (for spin chains in \cite{fc-11} and for Fermi gases in \cite{CMV-11b}).
For the Fermi gas this asymptotic result reads
\be
S_q =\frac1{12}\left(1+\frac1q\right)\ln\left (4N\sin\pi \frac\ell{L}\right)+ \frac{E_q}2 +o(N^0), 
\ee 
where $E_q$ is the same constant in Eq. (\ref{cnp}).
Also in this case, being the system homogeneous, the finite size scaling function can be entirely obtained from 
boundary conformal field theory \cite{CC-04,cc-rev}.

\section{Entanglement entropy for a quadratic trapping potential}
\label{sec3}

Let us now consider free fermions in an external harmonic potential (trap) 
\be
V(x) = \frac{1}{2} m \omega^2 x^2.
\ee
For simplicity in the following we set $\hbar=m=\omega=1$.
The dependence over the trap frequency $\omega$ can easily be restored using trap size scaling arguments \cite{CV-10}.
The single particle wave functions are
\be
\phi_n(x) = \frac{H_{n-1}(x)}{\sqrt{\pi^{1/2}  2^{n-1} (n-1)!}} e^{-x^2/2}\,, \quad n=1\dots N,
\ee
where $H_n(x)$ are the Hermite polynomials. The many body ground state wavefunction is
\be
\Psi_0(x_1,\dots x_N) = Z_N^{-1}  \prod_{i<j} (x_i-x_j) e^{- \sum_{i=1}^N x_i^2/2}, 
\ee 
with $Z_N$ a normalization constant.
Note that $|\Psi_0(x_1,\ldots, x_N)|^2$ can be interpreted as the joint
distribution of $N$ real eigenvalues $(x_1,\ldots, x_N)$ drawn from
the famous Gaussian unitary ensemble (GUE) \cite{Mehta}.
Using Christoffel-Darboux formula, the two-point function \eqref{cxy} is 
\be
C(x,y)= \frac{N^{1/2}}{\sqrt{2}} \frac{\phi_{N+1}(x) \phi_N(y)- \phi_{N}(x) \phi_{N+1}(y)}{x-y}\,,
\label{GUEkernel}
\ee
which is the well-known GUE kernel. 

The generating function for the particle number can be read from Eqs. (\ref{DA}) and (\ref{DA.2}) and it is
\bea \label{gener} 
\chi(s) \equiv \langle e^{-s N_A}\rangle= \det[\mathbb I + (e^{-s} -1) \mathbb A]\,,
\eea 
which, expanded to $O(s^2)$, yields the particle variance for an arbitrary subsystem $A$:
\be
V_{N_A}=\int_A dx \, C(x,x)-\int_A dx \int_A dy \, |C(x,y)|^2\,,
\ee
which is $V_{N_A}= {\rm Tr}[C_A-C_A^2]={\rm Tr}[\mathbb {A-A}^2]$.

For the harmonic potential, the entanglement entropy has been studied numerically in \cite{v-12,e-14} 
for several bipartition of the systems. 
The particle number variance has been studied numerically in the above manuscripts, but in the case when $A$
is a symmetric interval with respect to the centre of the trap of length $2\ell$, i.e. $A=[-\ell,\ell]$, random 
matrix theory allowed for a full large $N$ asymptotic analytical prediction for arbitrary value of $\ell$.
Three different scaling regimes have been identified  which are \cite{MMSV14}
\be
V_{N_A}\simeq \begin{cases}\displaystyle
 \frac{1}{\pi^{2}} \ln [N\zeta (2-\zeta^2)^{3/2}], &  \sqrt{2}-\zeta \sim O(1),
 \\
\\
       {\tilde V}_2( \sqrt{2}N^{2/3} (\zeta-\sqrt{2})), & \zeta-\sqrt{2} \sim O(N^{-\frac{2}3}),
\\ \\
      \exp[-2 N \phi(\zeta)], & \zeta-\sqrt{2} \sim O(1),
      \end{cases}
      \label{VNA}
\ee
where we introduced $\zeta= \ell/\sqrt{N}$ (notice that, in random matrix literature, lengths are always normalized to $\sqrt{N}$
as, e.g., in Ref. \cite{MMSV14}) and the functions
\bea
\tilde V_2(s)&=& 2\int_s^\infty K_{\rm Ai}(x,x)- 2\int_{[s,\infty]^2} \hspace{-4mm}  dx dy |K_{\rm Ai}(x,y)|^2, \nonumber \\
\phi(\zeta)&=& \frac{\zeta\sqrt{\zeta^2-2}}{2}+\ln \frac{\zeta-\sqrt{\zeta^2-2}}{\sqrt{2}},
\label{phizeta}
\eea
where $K_{\rm Ai}(x,y)$ is the Airy kernel
\be
K_{\rm Ai}(x,y)=\frac{{\rm Ai}(x){\rm Ai}'(y)- {\rm Ai}(y){\rm Ai}'(x)}{x-y}.
\label{KAi}
\ee 
We mention that while the scaling behavior of the variance in the intermediate edge regime in Eq. (\ref{VNA}) was well 
known \cite{s-00} (see also  \cite{e-14}), the full scaling function $\tilde V_2(s)$ (and in particular its asymptotic behaviors for both
negative and positive arguments) was computed explicitly only recently in \cite{MMSV14}.

In the following we will generalise the findings of Ref. \cite{MMSV14} to the entanglement entropy of a bipartite system 
in the case when $A$ is a symmetric interval around the centre of the trap.

\begin{figure}[t]
\includegraphics[width = \linewidth]{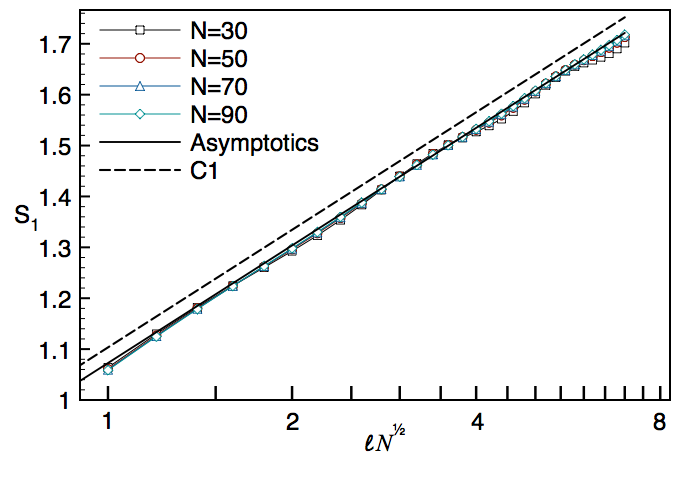}
\caption{(Color online)  Numerical evaluation of the entanglement entropy $S_1$ from the discretisation of Eq. (\ref{entaC}) 
for several values of $N$ and $\ell$ in the bulk regime $\ell\ll \sqrt{N}$. 
By increasing $N$ the data approach the asymptotic prediction (\ref{S.q.b}) in a non monotonic way. 
The dotted line is Eq. (\ref{entropy_bulk.2}) in which the additive constant has not been fixed to its correct value.}
\label{fig:Small}
\end{figure}

\subsection{Bulk regime: $ \zeta \sim 1/N \ll  1$}

We first consider the so-called bulk regime when $\zeta \sim 1/N  \ll  1$, i.e., when
the box size scales as the typical distance between eigenvalues of GUE
in the bulk, i.e., far away from the edges $\zeta =\sqrt{2}$ of the semi-circle. 
It is called bulk regime because the condition $\zeta\ll1$ ensures that the gas is 
almost homogeneous on these length scales. 

In this regime, when $N\to \infty$, $\zeta\to 0$ but keeping the product 
\bea
z= \frac{2\sqrt{2}\,N\zeta}{\pi},
\label{s_def}
\eea
fixed, it  has been proved~\cite{s-00,CL95,fs-95,fl-14} that the random
variable $N_A$ is indeed a {\em pure} Gaussian with mean
$\langle N_A\rangle\approx z$ and the variance 
\bea
V_{N_A}\approx V(z),
\label{variance.1}
\eea
where the scaling function $V(z)$ for all $z$ was first computed by Dyson 
and 
Mehta~\cite{DM62} and is given 
by (see e.g., appendix A.38 in Mehta's book~\cite{Mehta})
\bea
V(z)= z - 2 \int_0^z dr\, (z-r)\, \left[\frac{\sin(\pi r)}{\pi 
r}\right]^2\,.
\label{vs.1}
\eea
This function has the following asymptotics
\bea
V(z) &\to & z-\frac{1}{2}\, z^2 +O(z^3), \qquad\quad {\rm as}\quad z\to 0, 
 \\
& \to & \frac{1}{\pi^2} \ln (2\pi z) + \frac{(1+\gamma_E)}{\pi^2} + O(1/z), 
\quad {\rm as}\; z\to \infty,\nonumber
\label{vs_asymp}
\eea
where $\gamma_E=0.577215\dots$ is the Euler constant. 
Thus, in this range when $z\gg 1$, or equivalently $1/N\ll \zeta \ll 1$, the variance
behaves as
\bea
V_{N_A} = \frac{1}{\pi^2}\ln \left(2\sqrt{2}\, N\zeta \right)+ C_{\rm DM}+  O(1/z),
\label{variance.2}
\eea
where the constant $C_{\rm DM}$ is known as the Dyson-Mehta constant (see
A.38 in the book~\cite{Mehta}) and is given  by
\bea
C_{\rm DM}= \frac{(1+\gamma_E+\ln 2)}{\pi^2}=0.230036\dots
\label{dyson_mehta}
\eea

At this point, one would be tempted to use the fact that, in this bulk limit,
$N_A$ is a pure Gaussian and hence Eq. (\ref{entropy.6}) should be valid.
We anticipate that this is not the case, but before let us see what would be the prediction for the entropy 
under this assumption. In this case also $S_q$ becomes a function of the single scaling variable $z$ (cf. Eq. (\ref{s_def})) 
given by
\bea
S_q \stackrel{?}{=} \frac{\pi^2}{6}
\left(1+\frac{1}{q}\right)\, V(z)\,,
\label{entropy_bulk.1}
\eea
with $V(z)$ given in Eq. (\ref{vs.1}) for all $z$. In 
particular, for large $z$, i.e., when $\zeta \gg  1/N$ but still $\zeta \ll 1$,
using the large $z$ asymptotics of $V(z)$ in Eq. (\ref{vs_asymp}),  
one would get 
\bea
S_q \stackrel{?}{=} \frac{1}{6}\, \left(1+\frac{1}{q}\right)\, \ln 
\left(2\sqrt{2}\,N\zeta \right)+ C_q + \dots,
\label{entropy_bulk.2}
\eea
where the constant $C_q$ is 
\bea
C_q= \frac{\pi^2}{6}\, \Big(1+\frac{1}{q}\Big) C_{\rm DM}\,.
\label{c.1}
\eea
Notice the very simple dependence on $q$ of this constant compared with the 
fairly more complicated one in the case of homogeneous systems (cf. Eq. (\ref{cnp})).

The reasoning above has an obvious flaw.
Indeed, even if in the bulk regime the distribution of $N_A$ becomes Gaussian, by no means 
this implies that the full entropy is given by Eq. (\ref{entropy_bulk.2}):
the leading term in $N$ of the entropy is clearly correct, but non-Gaussian corrections to the distribution of $N_A$,
when integrated to calculate the entropy in Eq. (\ref{entropy.2}), can give rise to terms of the order $O(N^0)$ which add up to  
$C_q$ in Eq.  (\ref{entropy_bulk.2}). 
Indeed, these higher cumulants of $N_A$ have been calculated for a homogenous Fermi gas in \cite{CMV-12b} and their general
relation with the entropies have been studied in Refs. \cite{CMV-12b,KL-09,SRL-10,SRFKL-11}.

However, the subleading $O(N^0)$ term can be obtained by a general physical requirement. 
Indeed, close to the centre of the trap, the system is almost homogeneous with density $n(0)= N^{1/2} \sqrt{2}/\pi$.
Thus we expect the entanglement entropy to have the same value as a uniform system (cf. Eq. (\ref{FHres})) 
which for small $\ell$ is 
\be
S_q =\frac16\left(1+\frac1q\right)\ln\left (2 \frac{N}{L} \pi \ell\right)+ E_q
+\dots. 
\ee 
Replacing now the density $N/L$ with $n(0)= N^{1/2} \sqrt{2}/\pi$, we have the prediction
\be
S_q =\frac16\left(1+\frac1q\right)\ln\left (2\sqrt{2} N^{1/2}  \ell\right)+ E_q
+\dots, 
\label{S.q.b}
\ee 
which  has the same leading term as Eq. (\ref{entropy_bulk.2}), but presents a different additive constant. 
The two values $C_q$ and $E_q$ are indeed relatively close, for example at $q=1$ they are 
$C_1= 0.756788\dots $ and $E_1=0.726067\dots$.

In order to confirm the correctness of the previous reasoning, we compute numerically the entanglement entropy 
in this bulk regime. In Fig. \ref{fig:Small} we report the result for $q=1$ (but we checked also for other values of $q$).
It is evident that the data in this regime converges quickly (increasing $N$) to Eq. (\ref{S.q.b}).
It is also clear that changing the constant term from $E_1$ to $C_1$ moves the curve up 
of about $0.03$, which is a very visible shift on the vertical scale, as shown explicitly in Fig. \ref{fig:Small}. 
 
While the prediction in this bulk regime has been obtained on the sole basis of a scaling argument, 
this will  not be the case for the intermediate regime described in the following subsection.
However, having established the correct scaling behavior of the entanglement entropy in this regime, where the final 
result was known a priori, will be a very useful guide in the following subsection.

\subsection{Intermediate regime: $\zeta  \sim O(1)< \sqrt{2}- O(N^{2/3})$}

The question we answer in this subsection is what happens when one relaxes the upper limit
$\zeta \ll 1$, to $\zeta \sim O(1) < \sqrt{2}- O(N^{2/3})$, i.e., still far
from the edge scaling regime. 
In this regime, the full large deviation function associated with the distribution of $N_A$ was computed recently in \cite{MMSV14}
using a Coulomb gas method. From this large deviation function, the variance of $N_A$ can then be read off and it was 
found  to be a function of the single scaling variable \cite{MMSV14}
\bea
\Delta= N\,\zeta(2-\zeta^2)^{3/2}\,.
\label{delta_def}
\eea
The regime $\zeta \sim O(1) < \sqrt{2}- O(N^{2/3})$ translates
into the regime $\Delta\gg 1$
and it was shown recently~\cite{MMSV14} that the variance $V_{N_A}$ of $N_A$ behaves as
\bea
V_{N_A}= \frac{1}{\pi^2}\, \ln (\Delta) + C_{\rm DM} + O(1/\Delta)\,.
\label{var.2}
\eea
While the leading term was found analytically in Ref.~\cite{MMSV14}, the
subleading constant $C_{\rm DM}$ was found, by fitting numerical data, to
be the same as the Dyson-Mehta constant in Eq. (\ref{dyson_mehta}), see also \cite{v-12,Marino}.
Note that, in the limit $\zeta \ll \sqrt{2}$, using Eq. (\ref{delta_def}), the
result in Eq. (\ref{var.2}) reduces precisely to the bulk result in
Eq. (\ref{variance.2}), as it should.

\begin{figure}[t]
\includegraphics[width = \linewidth]{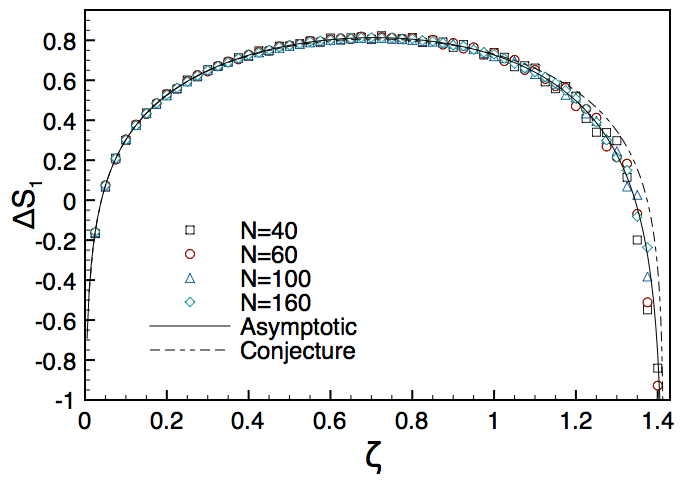}
\caption{(Color online)  Subtracted von Neumann entanglement entropy $\Delta S_1= S_1-(\ln N)/3$ as function of $\zeta=\ell/\sqrt{N}$ 
for several values of $N$ up to $N=160$. By increasing $N$ the data approach the asymptotic curve (\ref{entropy_whole.2}) 
in a non uniform way as function of $\zeta$. 
The dashed line is the conjecture in Eq. (\ref{entropy_whole.conj}) which is very close to the actual asymptotic curve 
everywhere except close to the edge. }
\label{fig:scal}
\end{figure}

The question is, can we use this result for the variance to compute
the entropy $S_q$. The main point is that the distribution of $N_A$
may no longer be a pure Gaussian and the entropy may have non-Gaussian corrections. 
Had the distribution been purely Gaussian with variance
$V_{N_A}$ given in Eq. (\ref{var.2}), we could use Eq. 
(\ref{entropy.6}) to obtain the prediction
\be
S_q\stackrel{?}{=} \frac{1}{6} \left(1+\frac{1}{q}\right) \ln
(N\,\zeta (2-\zeta^2)^{3/2})+ C_q + \dots,
\label{entropy_whole.1}
\ee
where the constant $C_q$ is given in Eq. (\ref{c.1}).
The prediction in Eq. (\ref{entropy_whole.1}) is valid assuming
$N_A$ is {\em purely} Gaussian with variance $V_{N_A}$ given in
Eq. (\ref{var.2}). 
However, the distribution of $N_A$ in this intermediate regime 
is not purely Gaussian and there are logarithmic 
corrections~\cite{MMSV14}. While, these logarithmic
corrections do not modify the leading term on the right hand side
of Eq. (\ref{entropy_whole.1}), they are expected to modify the subleading
$\zeta$-independent constant term $C_q$ (as in the bulk regime). 
However,  we can fix the constant term by requiring that, for small $\zeta$,
Eq. (\ref{entropy_whole.1}) reduces to the bulk one (\ref{S.q.b}), obtaining 
\be
S_q= \frac{1}{6} \left(1+\frac{1}{q}\right) \ln (N\,\zeta (2-\zeta^2)^{3/2})+ E_q +o(N^0) .
\label{entropy_whole.2}
\ee
This new prediction is one of the main results of this paper. 
Eq. (\ref{entropy_whole.2}) is indeed an expansion for $\Delta\gg 1$ of the 
scaling function for the entropy, in which $\Delta$ has been replaced with its actual 
value (\ref{delta_def}).

\begin{figure}[t]
\includegraphics[width = \linewidth]{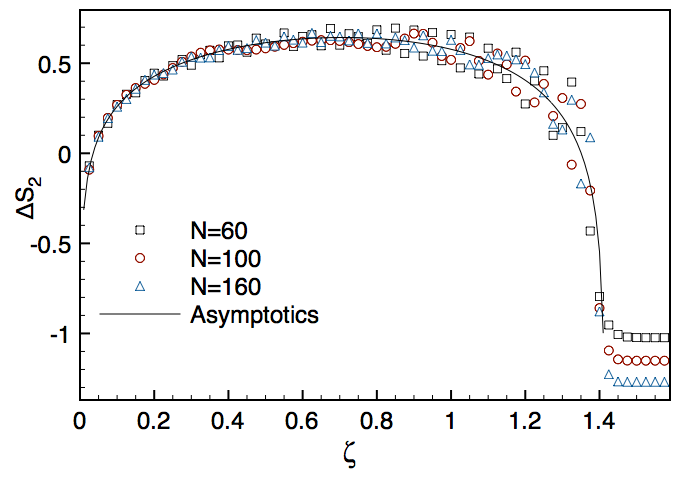}
\caption{(Color online)  Subtracted second order R\'enyi entropy $\Delta S_2= S_2-(\ln N)/4$ as function of $\zeta=\ell/\sqrt{N}$ 
for several values of $N$ up to $N=160$. 
The non-uniform approach to the asymptotic result (\ref{entropy_whole.2}) is more evident than in the case $q=1$.
}
\label{fig:R2}
\end{figure}

In Ref. \cite{v-12}, on the basis of the numerical data, it was conjectured that 
the R\'enyi entanglement entropies could have been described by the asymptotic form
\be
S_q \approx \frac{1}{6} \left(1+\frac{1}{q}\right) \ln
\Big (\frac{4 N}{\pi}  \sin \frac{\pi\zeta}{\sqrt2}\Big)+ E_q + \dots.
\label{entropy_whole.conj}
\ee
The two scaling curves are indeed very close to each other, but the numerical data for $q=1$
fit slightly better the random matrix prediction (\ref{entropy_whole.2}) compared to the above conjecture
(which however is very accurate, see Fig. \ref{fig:scal}). 
In Figs. \ref{fig:scal} and \ref{fig:R2} we report (for $q=1$ and $q=2$) the subtracted entropy
\be
\Delta S_q= S_q- \frac{1}{6} \left(1+\frac{1}{q}\right) \ln N,
\ee
which, in the limit of large $N$, is a scaling function of $\zeta=\ell/\sqrt{N}$.
Increasing $N$, the numerical data approach the random matrix prediction (\ref{entropy_whole.2}). 
For $q=1$ the agreement is very clear while for $q=2$ there are oscillating corrections to this asymptotic form 
(especially close to the edge) which make the distinction between Eq. (\ref{entropy_whole.2}) and 
the conjecture (\ref{entropy_whole.conj}) impossible.
As noticed already in Ref. \cite{v-12} the approach to the asymptotic result is non-uniform 
and gets very bad close to the edge, but, as we will show in the next subsection following Refs. \cite{MMSV14,e-14},
this apparently strange behavior can be understood in terms of the different scaling at the edge.

We have been also trying to describe, at least phenomenologically, the corrections to the asymptotic scaling behavior in the regime 
with $\Delta\gg1$ by subtracting to the numerical data the asymptotic prediction (\ref{entropy_whole.2}).
However, as it should be already clear from Fig. \ref{fig:R2} with $q=2$, at least two different kinds of corrections affects the data. 
The first is present also for small $\zeta$ in the form of small oscillations around the asymptotic value. 
This is reminiscent of the nowadays well understood ``unusual corrections'' to the scaling \cite{CCEN-10,ce-10,CC-10,ot-14,o-other}
which have been discussed in many different situations in homogeneous systems in which case they scale 
like $N^{-2/q}$ (for periodic systems). 
The second corrections instead originates from the edge $\ell\sim \sqrt{2N}$ and its form
will be derived in the next subsection.
However in the intermediate regime with $\zeta\sim O(1)$, a quantitative description of  
the corrections to the scaling eludes our understanding because  the two effects  are mixed up 
even for large, but finite, $N$.

\subsection{Edge regime}

\begin{figure}[t]
\includegraphics[width =\linewidth]{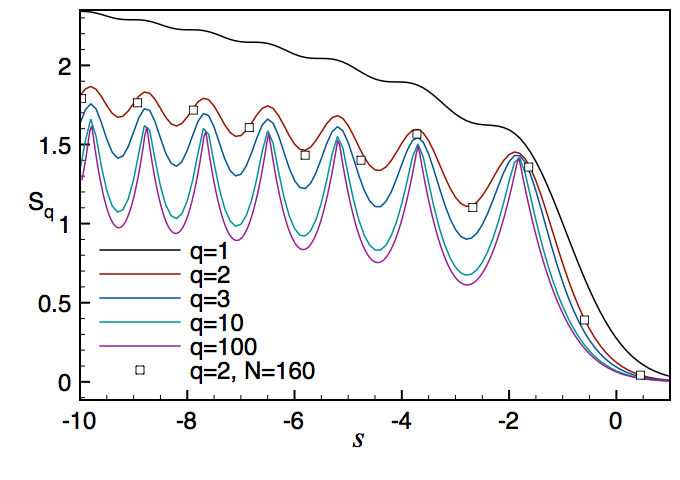}
\caption{(Color online)  Universal scaling of the R\'enyi entanglement entropy (always for $A=[-\ell,\ell]$ in a trapped gas) close to the edge.
We report the asymptotic curves in Eq. (\ref{S.edge}) as function of the scaling variable $s$ in Eq. (\ref{s:def}) for different 
values of $q$. We only report the numerical data for $N=160$ and $q=2$.}
\label{fig:edge}
\end{figure}

Close to the edge and in the limit of large $N$, the GUE kernel (\ref{GUEkernel})  tends to the Airy kernel (cf.  Eq. (\ref{KAi})) 
in terms of the scaling variable \cite{bb-91,f-93}
\be
s=\sqrt{2}N^{2/3} (\zeta-\sqrt{2}).
\label{s:def}
\ee
Since we are considering a symmetric interval with respect to the centre of the trap, 
there are two edges which contribute identically to the entanglement entropy. 
Thus, the large $N$ limit in the edge scaling regime is simply the limit of Eq. (\ref{entaC}), i.e. 
\footnote{One can indeed exchange $\tilde P_s K_{\rm Ai} \tilde P_s$ with $P_s K_{\rm Ai} P_s$ in this formula 
($\tilde P_s$ is the projection on the complementary interval) because entanglement entropies on
complementary sets are equal, i.e. Eq. (\ref{entaC}) is invariant under the replacement  $C_A \to C_{\bar A}$.
Furthermore, in the present scaling regime, for the contribution of each edge, one can neglect the presence of the other, resulting 
in the overall factor $2$ in Eq. (\ref{S.edge}). }
\be
S_q=\frac2{1-q}{\rm Tr} \ln [(P_s K_{\rm Ai} P_s)^q+(1-P_s K_{\rm Ai} P_s)^q], 
\label{S.edge}
\ee
where $P_s$ is the projector on the interval $[s,\infty]$.
This expression can be readily calculated from the spectrum of the operator 
$P_s K_{\rm Ai} P_s$, obtained by a proper discretisation following Ref. \cite{born}
(this procedure has been already applied  for $q=1$ in Ref. \cite{e-14}).
In Fig. \ref{fig:edge} we report the obtained exact scaling curve for $S_q$ as function of $s$ and for various values of $q$.
It is evident that the scaling curves present oscillations whose amplitude grows with increasing $q$.
This behavior explains why in the intermediate regime, the data for $S_1$ in Fig. \ref{fig:scal} are much better described by 
the asymptotic curve than the data for $S_2$ in Fig. \ref{fig:R2}.
The behavior of the amplitude of the oscillations is reminiscent of the one of the 
unusual corrections to the scaling \cite{CCEN-10,ce-10,CMV-11b}, but,  
being their origin different, if there is any connection between the two is still to be understood. 
Furthermore a similar behavior has been observed also close to the boundary of a hard-wall trap \cite{CMV-11b}, but in that case 
the theory of soft edge does not apply and the calculation of the asymptotic curve needs different methods.

Finally, we also checked that in the edge regime the numerical data approach the asymptotic result. 
This was already discussed in Ref. \cite{e-14} for $q=1$.
Thus in Fig. \ref{fig:edge} we limit to report a few data for $q=2$ and $N=160$. 
The agreement between the numerics and the prediction (\ref{S.edge}) is very good already for $N=160$.
We checked also other values of $q$, but we do not report them in order  to have a  readable figure.

\subsection{Beyond the edge}

There is clearly a third regime for $\zeta> \sqrt{2}$, in which the leading order of the entropy $S_q$ vanishes for large $N$. 
Thus $S_q$ is exponentially small. 
For sake of completeness, in this subsection we report the  calculation the entropy in this regime. 

In order to compute the leading correction we again start from Eq. (\ref{DA.2}). 
When $N_A=N$, i.e. all particles are inside the interval $[-\zeta,+\zeta[$, we have 
$D_A(\lambda)=(\lambda-1)^N$ and hence $\rho(a)=\delta(a-1)$. Consequently the
entropy is zero. The first elementary excitation has $N_A=N-1$,
which corresponds to pulling one fermion outside the Wigner sea (on either side). 
The probability of this event is given by \cite{edge} 
\be
p \sim e^{- 2 N \phi(\zeta)}\,, 
\ee
where $\phi(\zeta)$ is also the (right) large deviation 
function~\cite{review2014} 
associated with 
the largest eigenvalue of the
GUE random matrix given in Eq. (\ref{phizeta}). Note that as $\zeta \to \sqrt{2}$ from above, i.e. entering the small deviation (edge) regime, 
this function vanishes as
$\phi(\zeta) = \frac{2^{7/4}}{3} (\zeta-\sqrt{2})^{3/2}$, hence in this regime 
$p \sim e^{- \frac{4}{3} s^{3/2}}$ where $s$ is given in Eq. (\ref{s:def}). 

Since $p \ll 1$ in the regime $\zeta>2$ one has
\bea
{\rm Prob}(N_A) \approx p ~ \delta_{N_A,N-1} + (1-p) \delta_{N_A,N},
\eea 
which, using Eq. (\ref{DA.2}), leads to $D_A(\lambda)=(\lambda-1)^{N-1} (\lambda - (1-p))$ and hence 
\bea
\rho(a) \approx (1 - \frac{1}{N}) \delta(a-1) + \frac{1}{N} \delta(a-(1-p)). 
\eea
We thus obtain the entropy to leading order in small $p$
\bea
 S_q &=& \frac{N}{1-q} \int_0^1 da \ln(a^q + (1-a)^q) \rho(a) \\
& \simeq& \frac{1}{1-q} \ln\big( (1-p)^q + p^q \big) 
 \approx \frac{1}{1-q} \ln\big( 1 - q p  + p^q \big)  \nonumber.
\eea 
For fixed $q>1$ this is 
\bea \label{sqp}
S_q \simeq \frac{q}{q-1} p ,
\eea 
but on the other hand, for $q=1$, this reduces to
\bea \label{sqp1} 
S_1 \simeq - p \ln p .
\eea 
There exists a scaling function interpolating between the two limits.
Taking the limits $q-1 \to 0$ and $p \to 0$ keeping the product 
$y = - (q-1) \ln p$ fixed one has:
\bea
&& S_q(p) = (-  p \ln p) f(y), \quad f(y) = \frac{1-e^{-y}}{y} .
\eea 
Therefore for any $q \geq 1$ the entropy $S_q$ is exponentially small
in $N$ when $\zeta>\sqrt{2}$.

We can arrive at the same conclusion also starting from Eq.  
(\ref{S.edge}) which describes the behavior of the entropy in the small deviation 
regime from the edge.
In the region $\zeta >\sqrt{2}$ and $N$ large, the edge scaling variable
$s=\sqrt{2} N^{2/3} (\zeta - \sqrt{2})$ becomes large. Hence the operator
\begin{multline}
  \tilde K(x,y) \equiv [P_s K_{Ai} P_s](x,y) =\\
 = \theta(x) \theta(y) \int_0^{+\infty} dv  Ai(x+v+s) Ai(y+v+s) ,
\end{multline}
becomes uniformly small since $Ai(x) \sim e^{- \frac{2}{3} x^{3/2}}$ for large 
positive $x$. Hence we
can expand Eq. (\ref{S.edge}) as
\be
S_q \simeq  \frac{2}{1-q} {\rm Tr} \ln ( 1 - q \tilde K + \tilde K^q) .
\ee
For fixed $q>1$ and large $s$ we obtain
\bea
S_q = \frac{2 q}{1-q} {\rm Tr} \tilde K \sim \frac{q}{1-q} A(s) e^{- \frac{4}{3} s^{3/2}} ,
\eea 
where $A(s)$ is an unimportant prefactor which can easily be calculated
from the Airy function asymptotics. Thus it matches the result (\ref{sqp}) obtained
from the large deviation side. 
For $q=1$ one instead obtains
\bea
S_1 \simeq - 2 {\rm Tr}\, \tilde K \ln \tilde K ,
\eea 
which corresponds to Eq. (\ref{sqp1}). 

In conclusion both in the large deviation regime ($\zeta>\sqrt{2}$) as well as in the tail of the small deviation 
regime ($s \ll 1$) the entropy is exponentially small for large $N$. Here we limited to compute the
leading non vanishing term, but a systematic large $s$ expansion can in principle be performed
from Eq. (\ref{S.edge}). 

\section{Conclusions}
\label{conl}

In this manuscript we exploited and clarified the connection between entanglement entropy and random matrix theory 
for systems of free fermions.
Such a connection has been already (more or less explicitly) pointed out in the literature \cite{km-05,MMSV14,e-14},
but in this manuscript we push to the level to have a complete analytic description of the entanglement entropy in 
the ground-state of a free Fermi gas trapped by a harmonic potential.
The main analytical results of this  paper can be summarized by Eqs. (\ref{entropy_whole.2}) and (\ref{S.edge}).
Indeed, Eq. (\ref{entropy_whole.2}) provides the asymptotic behavior of the entropy in the scaling regime with $\ell/\sqrt{N}$
of order $1$, but far enough from the edge (a problem which was numerically studied in Ref. \cite{v-12}).
Instead Eq. (\ref{S.edge}) is the asymptotic behavior of the entropy in the edge scaling regime. 
Furthermore,  an interesting by-product of this work is that the 
entanglement entropy for finite number of particles (in some circumstances like the case of a trapped gas) 
can be more effectively calculated by 
ingeniously discretising the reduced correlation matrix (as described in Ref. \cite{born}) than by using the overlap matrix. 

We conclude by mentioning some possible extensions of this work which deserve further investigations. 
It would be interesting to understand whether random matrix theory could provide quantitative predictions 
not only for the ground state of a trapped Fermi gas, but also for excited states that in the homogeneous case 
present many interesting and universal features \cite{afc-09,abs-11,txas-13}.
Whether the present approach can be generalised to the entanglement entropy of free bosonic systems, 
such as the harmonic chain (see e.g. \cite{hc}), is also a relevant open question.  
Finally, generalisations to other entanglement estimators such as 
entanglement negativity \cite{neg}, entanglement contour \cite{cv-14}, or 
Shannon mutual information \cite{ar-13} are also waiting for an analytical description.

{\it Acknowledgements}. 
This work was initiated when all authors were guests of ICTP in Trieste, whose hospitality is kindly acknowledged. 
PC thanks Viktor Eisler for very fruitful correspondence.  
PC research was supported by ERC under the Starting Grant  n. 279391 EDEQS.
SNM acknowledges support from ANR grant 2011-BS04-013-01 WALKMAT and
from the Indo-French Centre for the Promotion of Advanced Research under Project 4604-3.
PLD acknowledges support from PSL grant ANR-10-IDEX-0001-02-PSL.

\appendix

\section{The distribution of eigenvalues of the overlap matrix}
\label{appa}

In this appendix, we report a technical by-product of this paper which is the distribution of eigenvalues of 
the overlap matrix (which is the same as the one of the reduced correlation matrix) for a trapped Fermi gas 
in the intermediate regime ($\zeta\sim O(1)$, but far from the edge).
At the leading order in $N$, for the interval $A=[-\ell,\ell]$, assuming the distribution of $N_A$ Gaussian, we have immediately 
\begin{multline}
D_A(\la)=\la^N \Big\langle  \Big(1 - \frac{1}{\la}\Big)^{N_A} \Big\rangle = \\
\la^N  e^{\langle N_A \rangle \ln(1 - \frac{1}{\la}) + \frac{\ln(N \zeta(2- \zeta^2)^{3/2})}{2 \pi^2} \ln^2(1 - \frac{1}{\la})},
\end{multline}
so that the resolvent function (\ref{Fla}) is 
\be
F(\la)=\frac{N}{\la} + \frac{\langle N_A\rangle}{\la(\la-1)}  + \frac{ \ln(N \zeta(2- \zeta^2)^{3/2})}{ \pi^2 \la (\la-1)} \ln(1 - \frac{1}{\la}).
\ee
The resulting distribution of eigenvalues $\rho(a)$, at the leading order in $N$, can be extracted from Eq. (\ref{rhoa}), giving 
\begin{multline}
\rho(a) = -\frac{1}{\pi} \lim_{\epsilon\to0^+} {\rm Im} F(a+i \epsilon) =\\
\Big(1-\frac{\langle N_A\rangle}{N}\Big ) \delta(a) +  \frac{\langle N_A\rangle}{N}\delta(a-1) +\\
+ \frac{ \ln(N \zeta(2- \zeta^2)^{3/2})}{N  \pi^2}  \frac{1}{a(1-a)} \,.
\label{rhoex}
\end{multline}
This distribution reproduces the correct leading order of the entropy.
Indeed by using 
\be
\frac{1}{1-q} \int_0^1 \frac{da}{a(1-a)}  \ln (a^q + (1-a)^q) = \frac{\pi^2}{6} \Big(1 + \frac{1}{q}\Big),
\ee
we obtain 
\begin{multline}
S_q =  \frac{N}{1-q} \int da\, \rho(a) \ln (a^q + (1-a)^q) = \\ \frac{ \ln(N \zeta(2- \zeta^2)^{3/2})}{ \pi^2} \frac{\pi^2}{6} \Big(1 + \frac{1}{q}\Big) , 
\label{ent.2}
\end{multline}
which coincides with the leading order of Eq. (\ref{entropy_whole.2}). 
Note that the third term in (\ref{rhoex}) actually is nonintegrable near $a=0$ and  $a=1$, however when the entropy is 
evaluated in Eq. (\ref{ent.2}), it gives a finite contribution.

%
%
%
%

\end{document}